\documentclass[twocolumn]{jpsj3}

\usepackage{color}
\usepackage{bm}

\title{%
Landau Levels on a Surface of Weak Topological Insulators
}

\author{%
Yositake Takane
}

\inst{%
Department of Quantum Matter, Graduate School of Advanced Sciences of Matter,\\
Hiroshima University, Higashihiroshima, Hiroshima 739-8530, Japan
}

\recdate{ \hspace{50mm} }

\abst{%
A three-dimensional weak topological insulator (WTI), being equivalent to
stacked layers of two-dimensional quantum spin-Hall insulators,
accommodates massless Dirac electrons on its side surface.
A notable feature of WTIs is that surface states typically consist of
two Dirac cones in the reciprocal space.
We study the Landau quantization of Dirac electrons of WTIs
in a perpendicular magnetic field.
It is shown that when the magnetic length $l_B$ is
much larger than the interlayer distance $a$,
surface electrons are quantized into Landau levels
according to the ordinary quantization rule for Dirac electrons.
It is also shown that, with decreasing $l_B$ toward $a$,
each Landau level and its spin state become modulated
in a nontrivial manner.
We demonstrate that this is attributed to the mixing of two Dirac cones
induced by the discreteness of the layered structure.
}

\begin{document}
\sloppy
\maketitle

\section{Introduction}

Three-dimensional (3D) weak topological insulators (WTIs)
can be regarded as stacked layers of
two-dimensional (2D) quantum spin-Hall (QSH) insulators.~\cite{fu,moore,roy}
The stacking direction is designated by
the weak vector $\mib{\nu} \equiv (\nu_1 \nu_2 \nu_3)$
with $\nu_1$, $\nu_2$, and $\nu_3$ being weak indices.
As a 2D QSH insulator possesses gapless excitations only at its edge
in the form of a one-dimensional (1D) helical channel,\cite{kane,bernevig}
a WTI accommodates low-energy electrons arising from
helical edge channels only on its side surface.
We refer to such surface electrons as Dirac electrons
since they obey the massless Dirac equation in the low-energy limit.
Notably, low-energy surface states of WTIs typically consist of
two Dirac cones in the reciprocal space.
Although disorder-induced scattering between two Dirac cones
is not forbidden, it does not necessarily extinguish
the topological nature of WTIs.~\cite{ran,imura1,ringel,mong,liu1,
imura2,yoshimura,kobayashi,morimoto1,obuse,takane}
Several materials have been proposed
as possible WTIs.~\cite{yan,rasche,tang,g-yang,pauly}

A characteristic feature of Dirac electrons shows up
when a perpendicular magnetic field is applied.
Dirac electrons are quantized into Landau levels
following the unique quantization rule that the energy of
the $n$th Landau level is proportional to $\sqrt{2n}$.~\cite{shon}
This has been experimentally observed in strong topological
insulators,~\cite{cheng,hanaguri} as in graphene.~\cite{novoselov}
Several authors have theoretically studied how the Landau levels are formed on
a surface of strong topological insulators.~\cite{liu2,z-yang,ilan,morimoto2}
However, an attempt similar to this has been lacking
for Dirac electrons in WTIs.
In this paper, we study the Landau quantization of Dirac electrons
on a side surface of WTIs in the presence of a perpendicular magnetic field.
To describe Dirac electrons in a magnetic field,
we mainly employ an effective 2D Hamiltonian composed of
coupled 1D chains of helical edge channels.~\cite{morimoto1,obuse,arita}
It has been demonstrated~\cite{arita} that this effective 2D model
can be derived from a 3D Wilson-Dirac Hamiltonian
for bulk topological insulators.~\cite{liu2}
We partly employ this 3D Hamiltonian to confirm the validity
of the 2D Hamiltonian.
The central question of this study is: does the Landau quantization in WTIs
reveal a unique behavior arising from the characteristic features
of the system under consideration?
The answer is yes.
If the side surface is infinitely long as we assume below, 
Landau levels are specified by the index $n$
and the wave number $k_{y}$ in the longitudinal direction.
In the weak magnetic field regime where the magnetic length $l_{B}$
is much larger than the interlayer distance $a$,
the behavior of the Landau level is basically explained
within the ordinary quantization rule for Dirac electrons.
However, a nontrivial behavior appears with decreasing $l_{B}$ toward $a$.
We find that, when $l_{B}/a$ becomes sufficiently small,
each Landau level and its spin state are modulated in an oscillatory manner
as a function of $l_{B}^{2}k_{y}/a$ with period 1
and the modulation becomes pronounced with increasing index $n$.
We demonstrate that this nontrivial behavior is attributed to the mixing of
two Dirac cones induced by the discreteness of the layered structure of WTIs.

In the next section, we present the effective 2D Hamiltonian as well as
the 3D Wilson-Dirac Hamiltonian,
and briefly explain the relationship between them.
In Sec.~3, we present the massless Dirac equation that
is derived from the 2D Hamiltonian in the continuum limit.
The quantization rule for Dirac electrons is derived on its basis.
In Sec.~4, we numerically determine the band structure of Dirac electrons
and calculate the spin expectation value of lower quantized levels
for several magnetic field strengths.
We find that each Landau level and its spin state become modulated
in a nontrivial manner with decreasing $l_B$ toward $a$.
In Sec.~5, we present an explanation for the nontrivial behavior.
Section~6 is devoted to a summary and discussion.
We set $\hbar = 1$ throughout this paper.

\section{Model}

%%%%%%%%%%%%%%%%%%
\begin{figure}[tbp]
\begin{center}
\includegraphics[height=3.0cm]{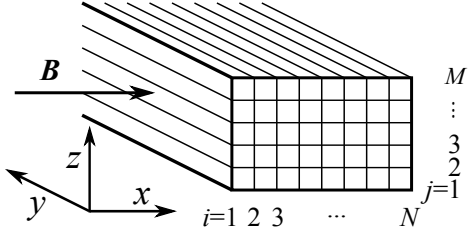}
\end{center}
\caption{Structure of a WTI sample considered in the text;
it consists of $M$ layers in the $z$-direction with $M \ge j \ge 1$
and $N$ sites in the $x$-direction with $N \ge i \ge 1$,
while it is infinitely long in the $y$-direction.
A magnetic field $\mib{B}$ is applied perpendicularly to the side surface
on the $yz$-plane.
}
\end{figure}
%%%%%%%%%%%%%%%%%%
Let us consider a WTI sample that consists of $M$ layers in the $z$-direction
and $N$ sites in the $x$-direction,
while it is infinitely long in the $y$-direction (see Fig.~1).
Let $j$ be the index used to specify the $M$ layers,
each of which is regarded as a 2D QSH insulator.
The $N$ sites in the $x$-direction are specified by the index $i$.
We assume that low-energy Dirac electrons emerge on the side surfaces
of height $M$ in the $yz$-plane at $i = 1$ and $N$.
If $N$ is chosen to be sufficiently large, the low-energy states
on one surface and those on the other surface are
mutually independent and exactly degenerate.
Thus, we hereafter consider only the side surface at $i = 1$.
Let $|j\rangle_{\uparrow}$ and $|j\rangle_{\downarrow}$
respectively be the basis vectors for right-going and left-going branches of
the 1D helical channel arising from the $j$th QSH layer.
Under the translational invariance in the $y$-direction,
surface electron states are characterized by the wave number $k_{y}$.
The effective 2D Hamiltonian for Dirac electrons on the side surface
is given by~\cite{morimoto1,obuse,arita}
\begin{align}
       \label{H_2D}
   H_{\rm 2D}
 & = \sum_{j=1}^{M}
     |j\rangle \left[ \begin{array}{cc}
                          vk_{y} & 0 \\
                          0 & -vk_{y}
                      \end{array} \right]
     \langle j|
   \nonumber \\
 &  \hspace{0mm}
   + \sum_{j=1}^{M-1}
     \left\{
     |j+1\rangle \left[ \begin{array}{cc}
                          0 & -\frac{1}{2a}v^{\prime} \\
                          \frac{1}{2a}v^{\prime} & 0
                          \end{array} \right]
     \langle j|
   + {\rm h.c.}
     \right\} ,
\end{align}
where $|j\rangle \equiv \left\{|j\rangle_{\uparrow},
|j\rangle_{\downarrow}\right\}$ and $a$ is the lattice constant,
which is simply the interlayer distance.
In this model, the Dirac points appear
at $(k_{y},k_{z}) = (0,0)$ and $(0,\pi/a)$.
For convenience, we define the $z$-coordinate of the $j$th chain as
\begin{align}
   z_{j} = \left(j-\frac{M+1}{2}\right)a
\end{align}
by setting the origin (i.e., $z = 0$) at the center of the system.
A perpendicular magnetic field $\mib{B} = (B, 0, 0)$
is introduced in terms of the vector potential $\mib{A} = (0, A_y,0)$ with
\begin{align}
     \label{eq:def-A}
   A_y = -Bz .
\end{align}
With this choice of $\mib{A}$, the wave number $k_{y}$ remains
a good quantum number since the translational invariance in the $y$-direction
is preserved.
The effect of $\mib{B}$ can be taken into account by replacing $k_{y}$
in the Hamiltonian with $k_{y} + eA_{y}$.
The Zeeman term is ignored in the main part of our analysis,
and its effect is briefly discussed in Sec.~6.

To supplement $H_{\rm 2D}$, we also use the 3D Wilson-Dirac Hamiltonian
for topological insulators:~\cite{liu2}
\begin{align}
   H_{\rm 3D}
   = \left[ 
       \begin{array}{cccc}
         M_{\mib{k}} & A_{\perp}k_{z}a & 0 & A_{\parallel}k_{-}a \\
         A_{\perp}k_{z}a & -M_{\mib{k}} & A_{\parallel}k_{-}a & 0 \\
         0 & A_{\parallel}k_{+}a & M_{\mib{k}} & -A_{\perp}k_{z}a \\
         A_{\parallel}k_{+}a & 0 & -A_{\perp}k_{z}a & -M_{\mib{k}}
       \end{array}
     \right] ,
\end{align}
where $k_{\pm} = k_{x}\pm i k_{y}$ and
\begin{align}
  M_{\mib{k}}
    = m_0 + m_{2\parallel}(k_{x}^{2}+k_{y}^{2})a^{2}
          + m_{2\perp}k_{z}^{2}a^{2} .
\end{align}
We discretize the $x$- and $z$-coordinates and implement $H_{\rm 3D}$
on the square lattice in the $xz$-plane
leaving the $y$-coordinate unchanged.~\cite{arita}
In accordance with our assumption that the system is equivalent to
2D QSH insulators stacked in the $z$-direction,
we focus on the weak topological phase with $\mib{\nu} \equiv (001)$.
After the discretization, the Wilson mass term $M_{\mib{k}}$ is modified to
\begin{align}
        \label{eq:Mk-dis}
  M_{\mib{k}}^{\rm dis}
  & = m_0 + m_{2\parallel}\left\{2[1-\cos(k_{x}a)]+(k_{y}a)^{2}\right\}
      \nonumber \\
  & \hspace{5mm}
    + m_{2\perp}2[1-\cos(k_{z}a)] .
\end{align}
The weak topological phase with $\mib{\nu} \equiv (001)$ is
stabilized when the parameters satisfy~\cite{imura2}
\begin{align}
    \label{eq:condition-WTI}
  m_{2\parallel} > \frac{1}{4}|m_0| > m_{2\perp}
                > \frac{1}{4}|m_0| - m_{2\parallel} ,
\end{align}
where $m_{2\parallel} > 0 > m_{0}$ is assumed.
This condition fixes the sign of the mass term at four symmetric points
on the $k_{x}k_{z}$-plane with $k_{y}=0$ as follows:
\begin{align}
        \label{eq:Mk-dis><0}
  M_{\mib{k}}^{\rm dis}
  = \left\{ \begin{array}{ll}
               m_{0}<0,
                & \mib{k}' = (0,0) \\
               m_{0}+4m_{2\parallel}>0,
                & \mib{k}' = (\pi/a,0) \\
               m_{0}+4m_{2\perp}<0,
                & \mib{k}' = (0,\pi/a) \\
               m_{0}+4m_{2\parallel}+4m_{2\perp}>0,
                & \mib{k}' = (\pi/a,\pi/a)
            \end{array}
    \right. , 
\end{align}
where $\mib{k}' = (k_{x},k_{z})$.
This indicates that, on the side surface in the $yz$-plane,
the Dirac point appears at $(k_{y},k_{z}) = (0,0)$ and $(0,\pi/a)$
in accordance with the 2D model.
Again, the effect of $\mib{B}$ can be taken into account
by the replacement of $k_{y}$ with $k_{y} + eA_{y}$.

It should be pointed out that
the 2D model can be derived from the 3D model.~\cite{arita}
The velocities in the 2D model are directly related to the parameters
of the 3D model,
\begin{align}
  v & = A_{\parallel}a ,
       \\
  v^{\prime} & = A_{\perp}a .
\end{align}

\section{Continuum Dirac Theory}

Let us consider the effective 2D Hamiltonian
given in Eq.~(\ref{H_2D}) in the continuum limit.
Note that our model involves the two Dirac cones.
Hereafter, the Dirac cones centered at $(k_{y},k_{z})=(0, 0)$ and $(0, \pi/a)$
are respectively referred to as the first and second Dirac cones.
In the continuum limit, the effective Hamiltonians $H_{+}$ and $H_{-}$
describing low-energy states in the first and second Dirac cones
are given by
\begin{align}
   H_{\pm}
   = \left[ 
       \begin{array}{cc}
         v(k_{y}-eBz) & \mp v^{\prime}\partial_{z} \\
         \pm v^{\prime}\partial_{z} & -v(k_{y}-eBz)
       \end{array}
     \right] ,
\end{align}
where the vector potential is explicitly included.
It is convenient to parameterize the strength of $B$
in terms of the magnetic length defined by
\begin{align}
   l_B = \frac{1}{\sqrt{eB}} .
\end{align}
The anisotropy of the system is characterized by
\begin{align}
   r = \frac{v^{\prime}}{v} .
\end{align}

We determine the eigenstates of $H_{\pm}$ by using
the annihilation and creation operators satisfying
the commutation relation of $[a, a^{\dagger}] = 1$, defined by
\begin{align}
  a & = \frac{\tilde{l}_{B}}{\sqrt{2}}
        \left[\frac{1}{\tilde{l}_{B}^{2}}(z-z_{\rm c}(k_{y}))
              + \partial_{z}  \right] ,
          \\
  a^{\dagger}
    & = \frac{\tilde{l}_{B}}{\sqrt{2}}
        \left[\frac{1}{\tilde{l}_{B}^{2}}(z-z_{\rm c}(k_{y}))
              - \partial_{z}  \right] ,
\end{align}
where $\tilde{l}_{B} = \sqrt{r}l_{B}$ and
$z_{\rm c}(k_{y}) = l_{B}^{2}k_{y}$.
With these operators, $H_{\pm}$ is rewritten as
\begin{align}
   H_{\pm}
   = \frac{\sqrt{r}v}{\sqrt{2}l_{B}}
     \left[ 
       \begin{array}{cc}
         -\left(a+a^{\dagger}\right) & \mp\left(a-a^{\dagger}\right) \\
         \pm\left(a-a^{\dagger}\right) & \left(a+a^{\dagger}\right)
       \end{array}
     \right] .
\end{align}
The eigenvalues are quantized into the Landau levels as
\begin{align}
  E_{\pm n} = \pm \frac{\sqrt{r}v}{l_B}\sqrt{2n}
\end{align}
with $n \ge 0$.
The wave functions are expressed in terms of the eigenstates of
the number operator $a^{\dagger}a$ satisfying
\begin{align}
   a^{\dagger}a \psi_n(z) = n \psi_n(z) ,
\end{align}
where the explicit forms of the lower three functions are given by
\begin{align}
   \psi_{0}(z)
  & = \frac{1}{\sqrt{\pi\tilde{l}_{B}}}
      e^{-\frac{(z-z_{\rm c}(k_{y}))^{2}}{2\tilde{l}_{B}^{2}}} ,
        \\
   \psi_{1}(z)
  & = \frac{\sqrt{2}}{\sqrt{\pi\tilde{l}_{B}}}
      \left(\frac{z-z_{\rm c}(k_{y})}{\tilde{l}_{B}}\right)
      e^{-\frac{(z-z_{\rm c}(k_{y}))^{2}}{2\tilde{l}_{B}^{2}}} ,
        \\
   \psi_{2}(z)
  & = \frac{\sqrt{2}}{\sqrt{\pi\tilde{l}_{B}}}
      \left[ \left(\frac{z-z_{\rm c}(k_{y})}{\tilde{l}_{B}}\right)^{2}
                   - \frac{1}{2} \right]
      e^{-\frac{(z-z_{\rm c}(k_{y}))^{2}}{2\tilde{l}_{B}^{2}}} .
\end{align}
Note that $\psi_n(z)$ depends on $k_{y}$
since it is centered at $z_{\rm c}(k_{y})$.
The eigenstate of $H_{+}$ with the energy $E_{\pm n}$ is expressed as
\begin{align}
   \mib{\phi}^{+}_{\pm n}(z)
   = \frac{1}{2}
     \left[ 
        \begin{array}{c}
           \psi_{n}(z) \mp \psi_{n-1}(z) \\
           \psi_{n}(z) \pm \psi_{n-1}(z)
        \end{array}
     \right]
\end{align}
for $n > 0$ and
\begin{align}
   \mib{\phi}^{+}_{0}(z)
   = \frac{1}{\sqrt{2}}
     \left[ 
        \begin{array}{c}
           \psi_{0}(z) \\
           \psi_{0}(z)
        \end{array}
     \right] .
\end{align}
The eigenstate of $H_{-}$ with the energy $E_{\pm n}$ is expressed as
\begin{align}
   \mib{\phi}^{-}_{\pm n}(z)
   = \frac{1}{2}
     \left[ 
        \begin{array}{c}
           \psi_{n}(z) \mp \psi_{n-1}(z) \\
           -\psi_{n}(z) \mp \psi_{n-1}(z)
        \end{array}
     \right]
\end{align}
for $n > 0$ and
\begin{align}
   \mib{\phi}^{-}_{0}(z)
   = \frac{1}{\sqrt{2}}
     \left[ 
        \begin{array}{c}
           \psi_{0}(z) \\
           -\psi_{0}(z)
        \end{array}
     \right] .
\end{align}
The spin state of $\mib{\phi}^{\pm}_{m}$ with $m = 0, \pm 1, \pm 2, \cdots$
is determined by the expectation value of the Pauli spin matrices,
\begin{align}
  \langle \phi^{\pm}_{m} \left| \mib{\sigma}
  \right| \phi^{\pm}_{m} \rangle
  = \int dz \left[\mib{\phi}^{\pm}_{m}(z)\right]^{\dagger}
            \mib{\sigma} \mib{\phi}^{\pm}_{m}(z) .
\end{align}
We easily obtain
\begin{align}
  \langle \phi^{\pm}_{m} \left|
  \mib{\sigma} \right| \phi^{\pm}_{m} \rangle
  = (0, 0, 0)
\end{align}
for $m \neq 0$ and
\begin{align}
  \langle \phi^{\pm}_{0} \left|
  \mib{\sigma} \right| \phi^{\pm}_{0} \rangle
  = (\pm 1, 0, 0) .
\end{align}

The above argument provides the quantization rule for Dirac electrons in WTIs.
Firstly, the energy of the $n$th Landau level is quantized into a value
proportional to $\sqrt{2n}$, as in monolayer graphene.~\cite{shon,novoselov}
Secondly, each Landau level is doubly degenerate,
reflecting the existence of the two Dirac cones.
Thirdly, the Landau levels are not spin-polarized
except for the $0$th Landau level $\mib{\phi}^{\pm}_{0}(z)$,
which is polarized to the $\pm x$-direction.
This rule is reliable when $l_{B}/a \gg 1$.

\section{Numerical Results}

%%%%%%%%%%%%%%%%%%
\begin{figure}[tbp]
\begin{center}
\includegraphics[height=5.0cm]{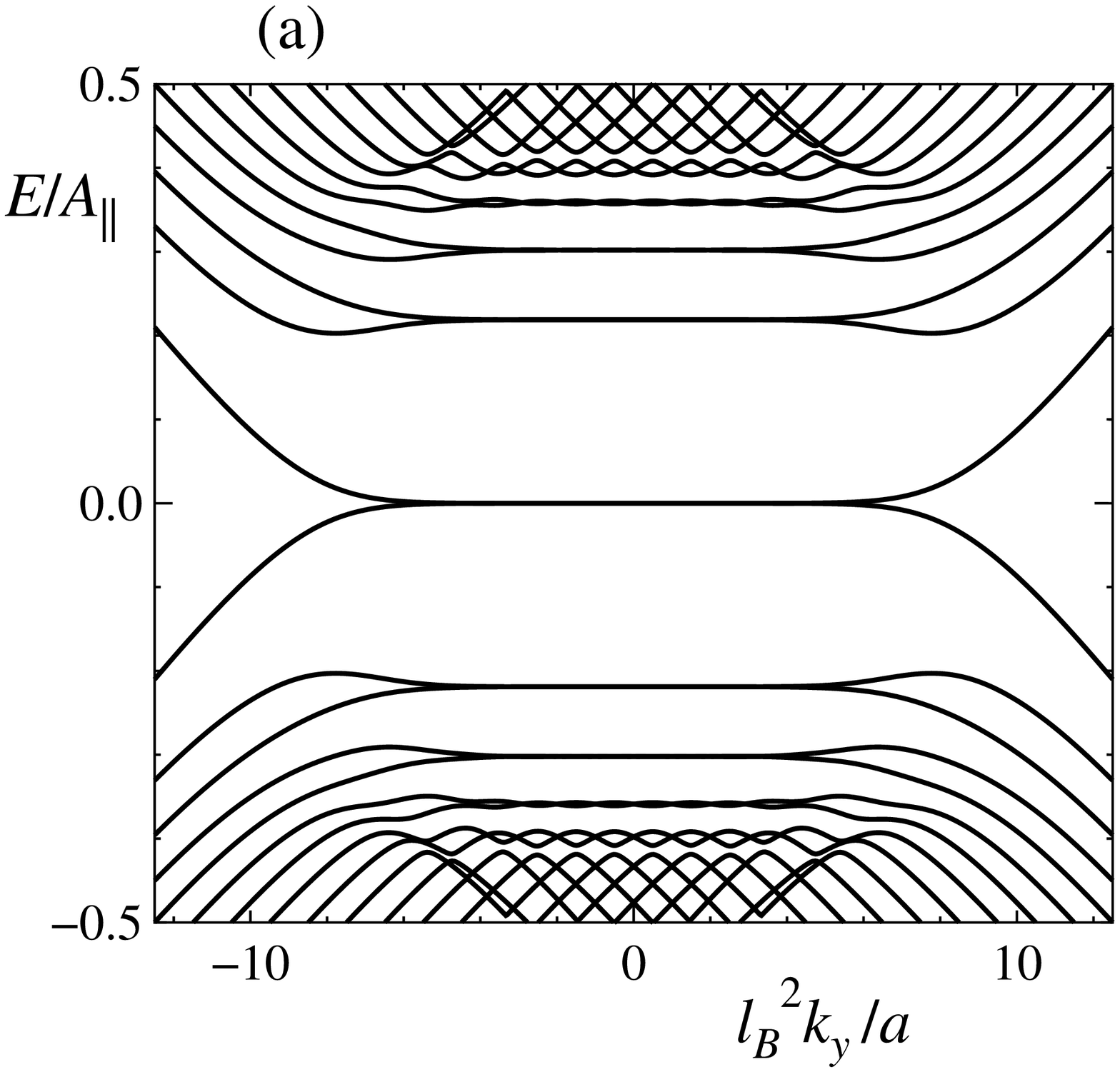}
\includegraphics[height=5.0cm]{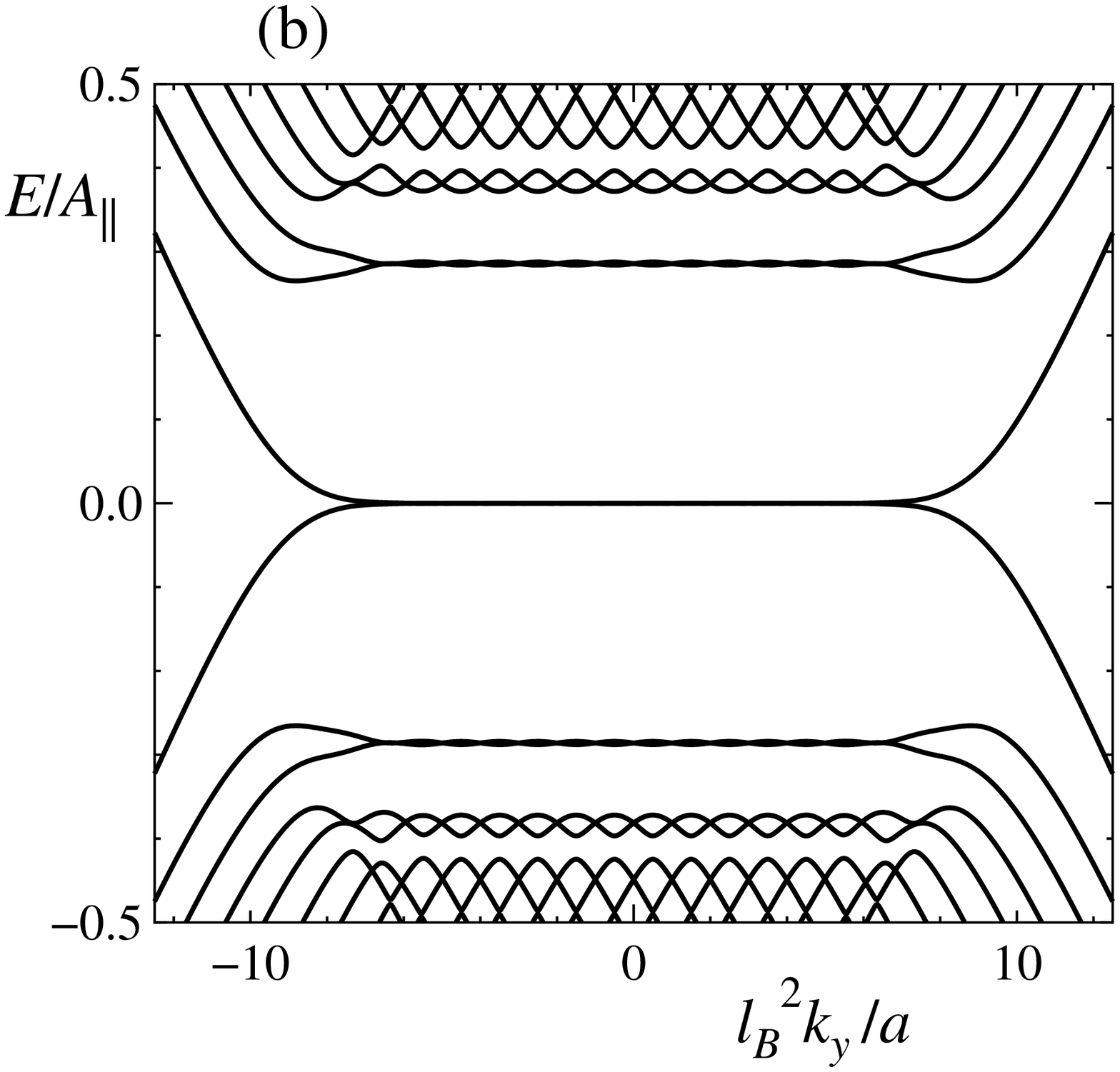}
\includegraphics[height=5.0cm]{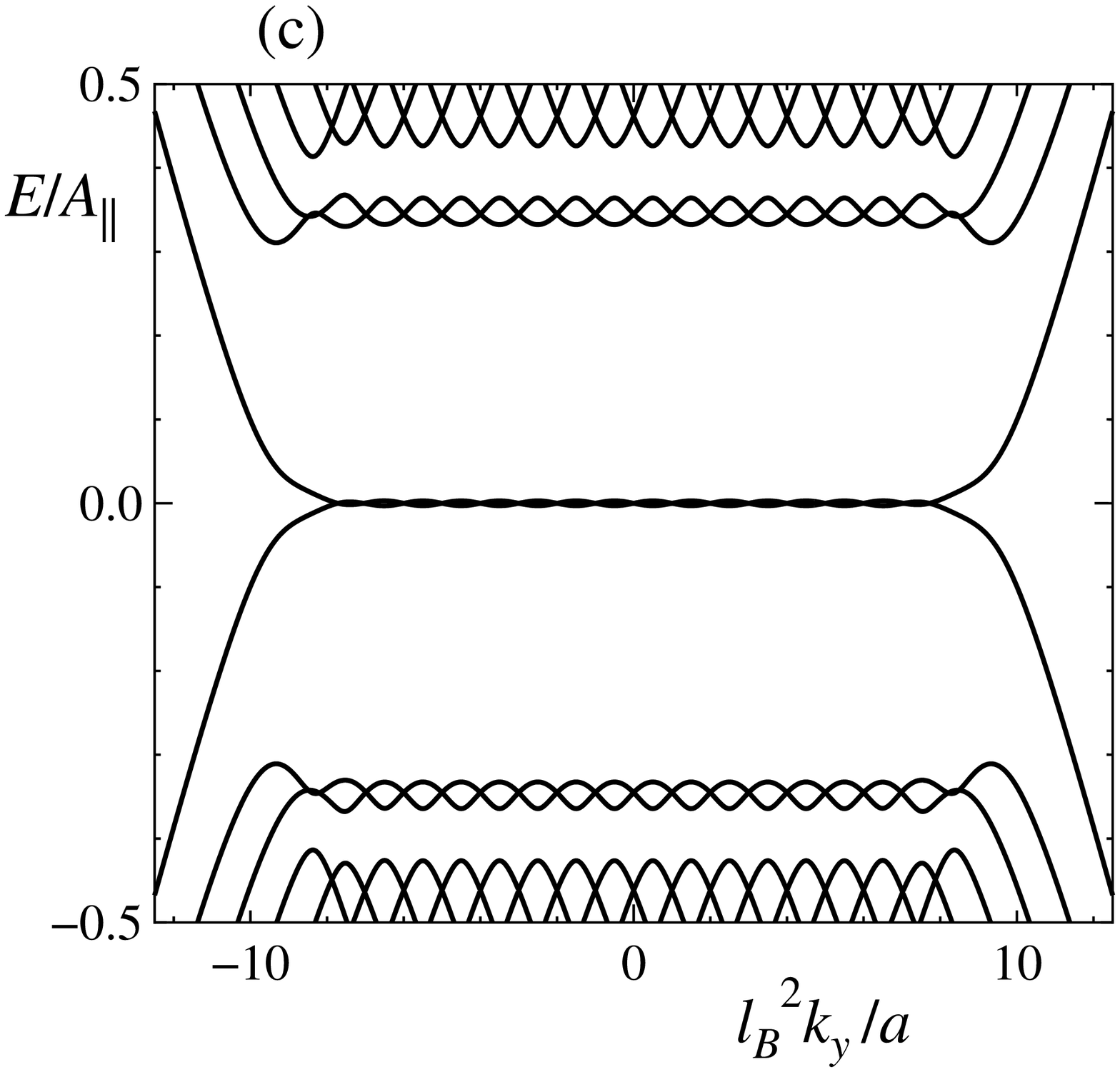}
\end{center}
\caption{
Band structures in the cases of
(a) $l_{B}/a = 4$, (b) $l_{B}/a = 3$, and (c) $l_{B}/a = 2.4$
with $r = 0.4$ and $M = 21$.
}
\end{figure}
%%%%%%%%%%%%%%%%%%
We numerically determine the band structure of Dirac electrons
in the presence of a perpendicular magnetic field
on the basis of the effective 2D Hamiltonian $H_{\rm 2D}$.
We also calculate the expectation values of $z$ and $\mib{\sigma}$ to observe
whether the quantization rule obtained from the continuum Dirac theory
is applicable even in a strong magnetic field.

Figure~2 shows the band structure of the system with
$r \equiv v^{\prime}/v = 0.4$ and $M = 21$.
We see that several dispersion-less flat subbands
appear in a restricted region of $k_{y}$.
They correspond to Landau levels.
Inside the displayed energy range of $|E/A_{||}| \le 0.5$,
we can identify the Landau levels with the indices $n = 0$, $\pm 1$, $\pm 2$,
$\pm 3$, and $\pm 4$ in the weak magnetic field case of $l_{B}/a = 4$.
In the stronger case of $l_{B}/a = 3$,
the Landau levels with $n = \pm 3$ and $\pm 4$ cannot be identified.
Note that our system is restricted to the finite width of $(M-1)a/2 \ge |z|$.
Thus, roughly speaking, the Landau levels with a small $n$ stably exist
under the condition of $(M-1)/2-l_{B}/a \gtrsim l_{B}^{2}|k_{y}|/a$
as $\psi_{n}(z)$ is centered at $z_{\rm c} = l_{B}^{2}k_{y}$ and
its spatial range is on the order of $l_{B}$.~\cite{comment}
In the case of $M = 21$ and $l_{B}/a = 4$, the condition is simplified to
$6 \gtrsim l_{B}^{2}|k_{y}|/a$,
in good agreement with the result shown in Fig.~2.
Outside of this region,
each energy level shows a dispersion as a function of $k_{y}$.
This suggests the appearance of chiral edge states.
We also see that each Landau level is doubly degenerate,
reflecting the existence of the two Dirac cones.
A peculiar feature of the Landau levels shown in Fig.~2 is that
they deviate from the flat dispersion and are modulated
in an oscillatory manner as a function of
$l_{B}^{2}k_{y}/a$ with a period of 1.
This becomes pronounced with
decreasing $l_{B}$ and/or increasing the index $n$.
The oscillatory behavior cannot be explained
within the continuum Dirac theory.

%%%%%%%%%%%%%%%%%%
\begin{figure}[tbp]
\begin{center}
\includegraphics[height=4.5cm]{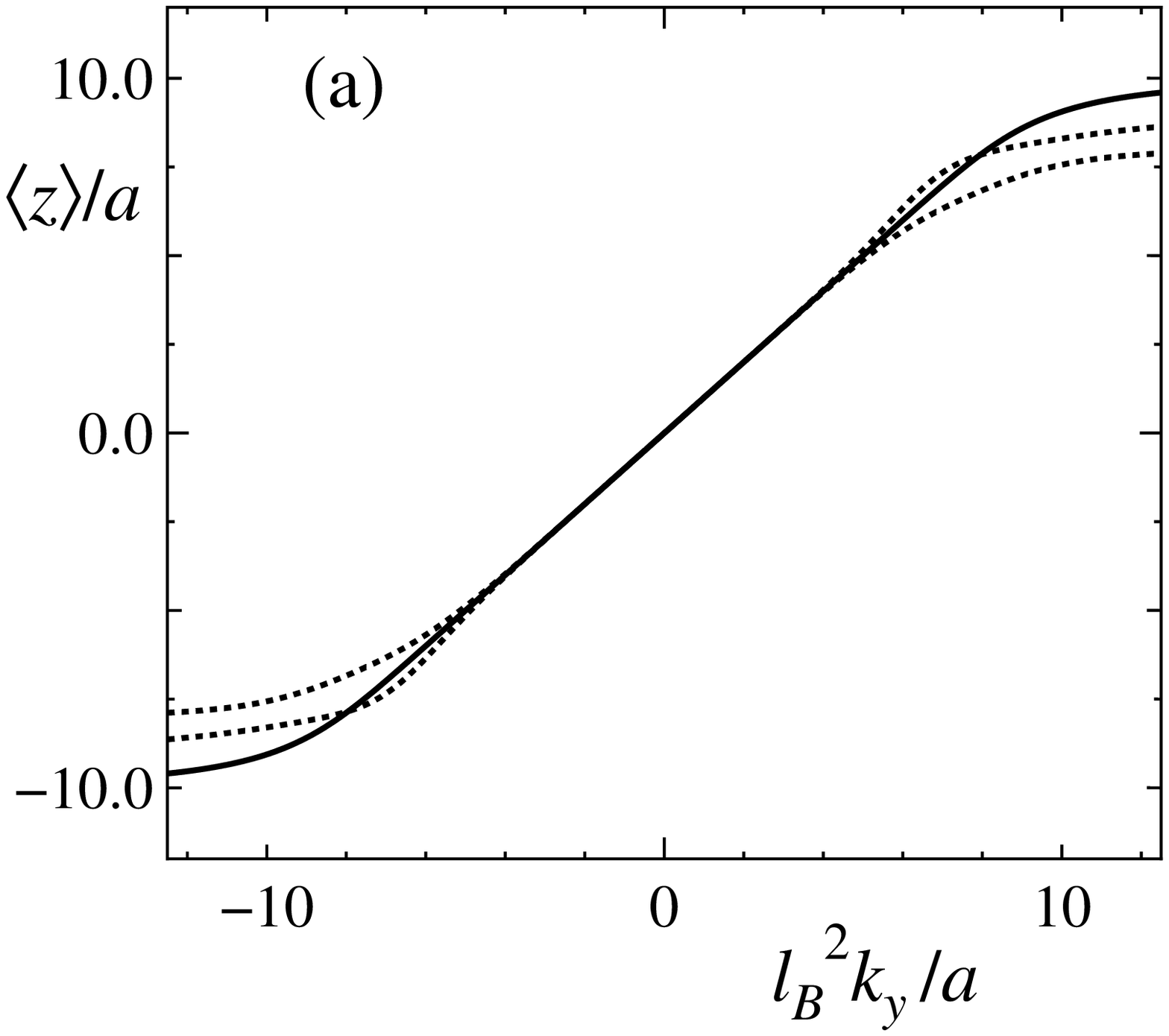}
\includegraphics[height=4.5cm]{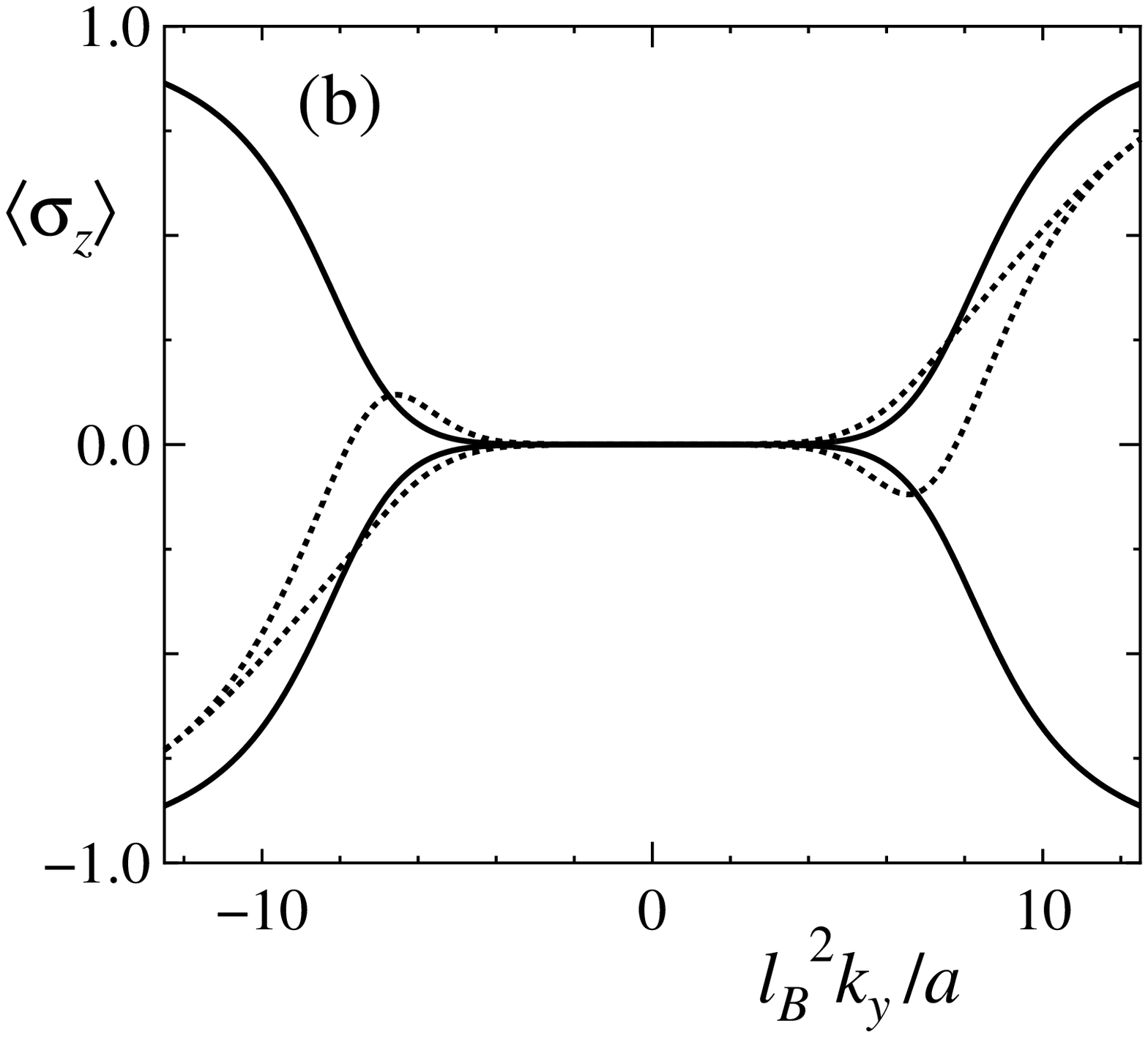}
\end{center}
\caption{
Expectation values in the case of $l_{B}/a = 4$ with $r = 0.4$ and $M = 21$:
(a) $\langle z\rangle/a$ and (b) $\langle \sigma_{z} \rangle$
in the $0$th Landau level (solid line)
and in the $1$st Landau level (dotted line).
}
\end{figure}
%%%%%%%%%%%%%%%%%%
%%%%%%%%%%%%%%%%%%
\begin{figure}[tbp]
\begin{center}
\includegraphics[height=4.5cm]{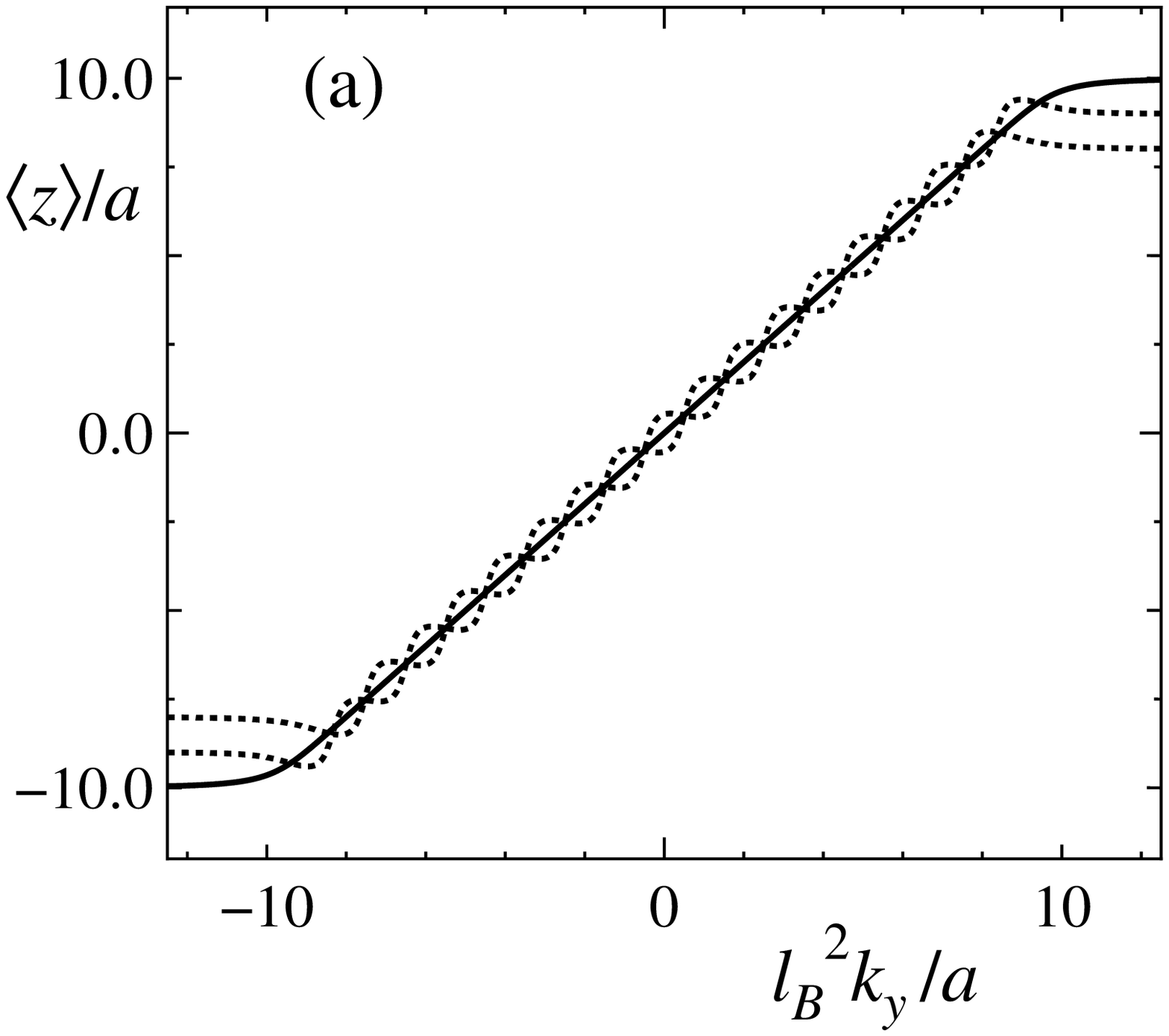}
\includegraphics[height=4.5cm]{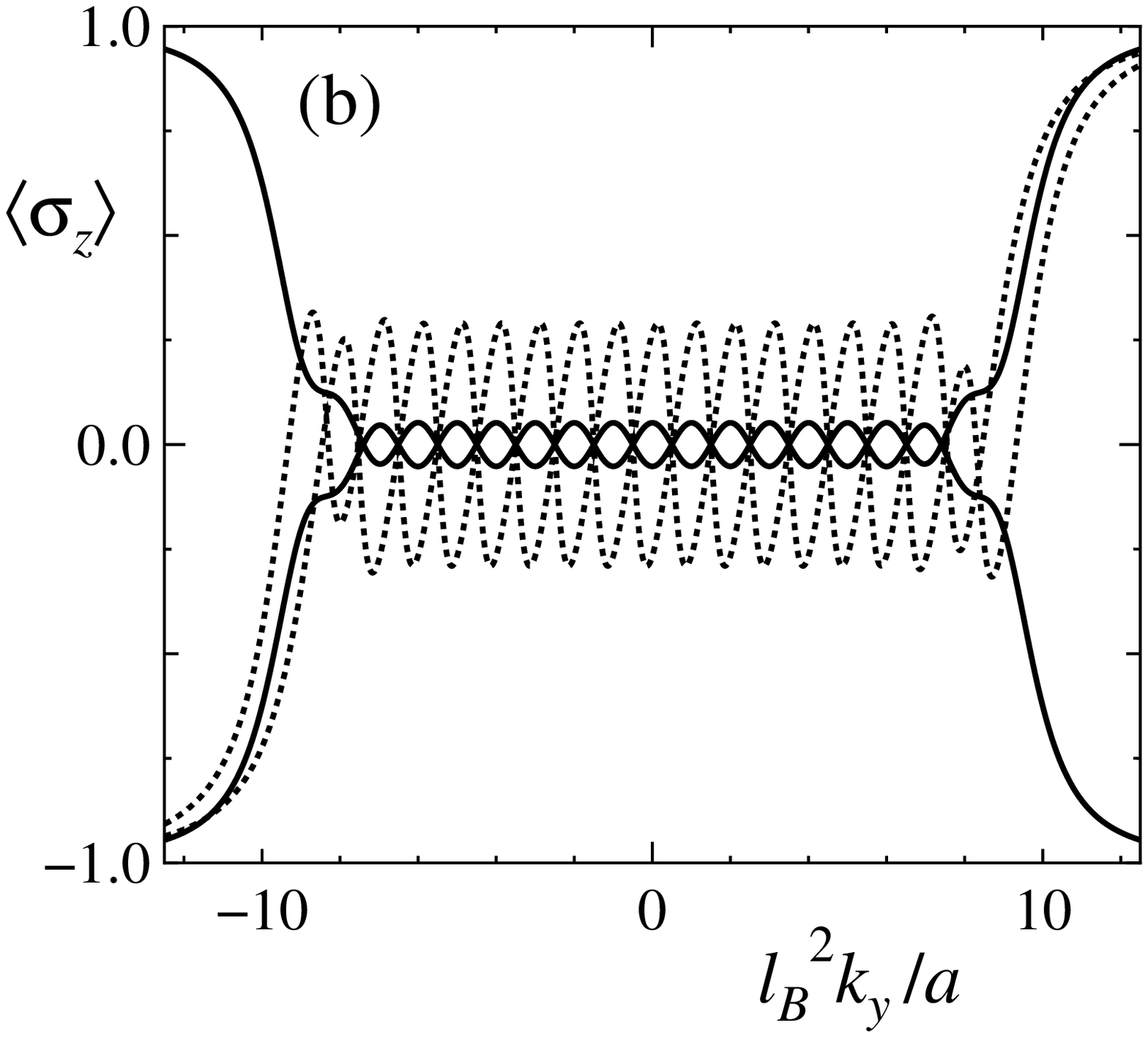}
\end{center}
\caption{
Expectation values in the case of $l_{B}/a = 2.4$ with $r = 0.4$ and $M = 21$:
(a) $\langle z\rangle/a$ and (b) $\langle \sigma_{z} \rangle$
in the $0$th Landau level (solid line)
and in the $1$st Landau level (dotted line).
}
\end{figure}
%%%%%%%%%%%%%%%%%%
Figure~3 shows the expectation values of $z$ and $\sigma_z$
in the $0$th and $1$st Landau levels in the case of $l_{B}/a = 4$
with $r = 0.4$ and $M = 21$.
We see that $\langle z \rangle/a$ is proportional to $k_{y}$
in the central region of $6 \gtrsim l_{B}^{2}|k_{y}|/a$.
Outside of this region, $\langle z \rangle/a$ approaches
the edge position (i.e., $\pm 10$) with increasing $|k_{y}|$,
indicating the appearance of chiral edge states.
This is also confirmed by the result of $\langle \sigma_z \rangle$
in the region of $l_{B}^{2}|k_{y}|/a \gtrsim 10$,
indicating that the spin is polarized to the $\pm z$-direction
(i.e., $\langle \sigma_z \rangle \sim \pm 1$) with increasing $|k_{y}|$.
As an edge state going to the positive (negative) direction is mainly
composed of spin-up (spin-down) channels
near the corresponding edge of the system,
its spin should be polarized to the $z$-direction ($-z$-direction).
In the central region of $6 \gtrsim l_{B}^{2}|k_{y}|/a$,
$\langle \sigma_z \rangle$ vanishes in accordance with
the continuum Dirac theory.
The expectation value of $\sigma_{x}$ as well as that of $\sigma_{y}$ vanishes
not only in the $1$st Landau level but also in the $0$th Landau level
regardless of $k_{y}$ (data not shown).
This is in contrast to the prediction of the continuum Dirac theory
(i.e., the $0$th Landau level is spin-polarized to the $\pm x$-direction),
implying that the corresponding states in the first and second Dirac cones
are hybridized to form spin-unpolarized states.

Figure~4 shows the expectation values of $z$ and $\sigma_z$
in the $0$th and $1$st Landau levels in the case of $l_{B}/a = 2.4$
with $r = 0.4$ and $M = 21$.
We see that $\langle \sigma_z \rangle$ oscillates as a function of
$l_{B}^{2}k_{y}/a$ with a period of 1.
In the $1$st Landau level, $\langle z \rangle/a$ also shows
an oscillatory behavior.
These features cannot be explained within the continuum Dirac theory.

%%%%%%%%%%%%%%%%%%
\begin{figure}[btp]
\begin{center}
\includegraphics[height=5.0cm]{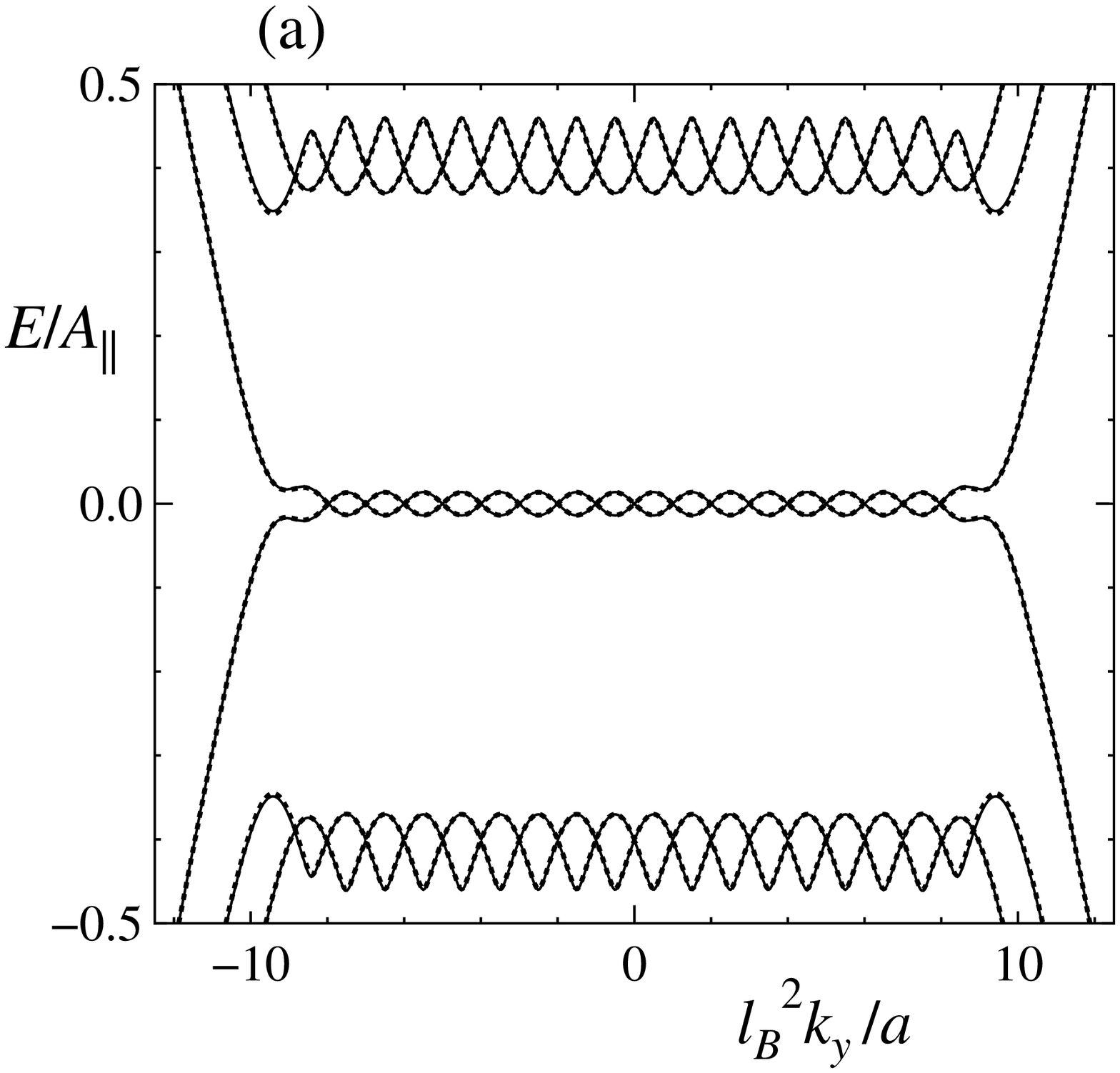}
\includegraphics[height=5.0cm]{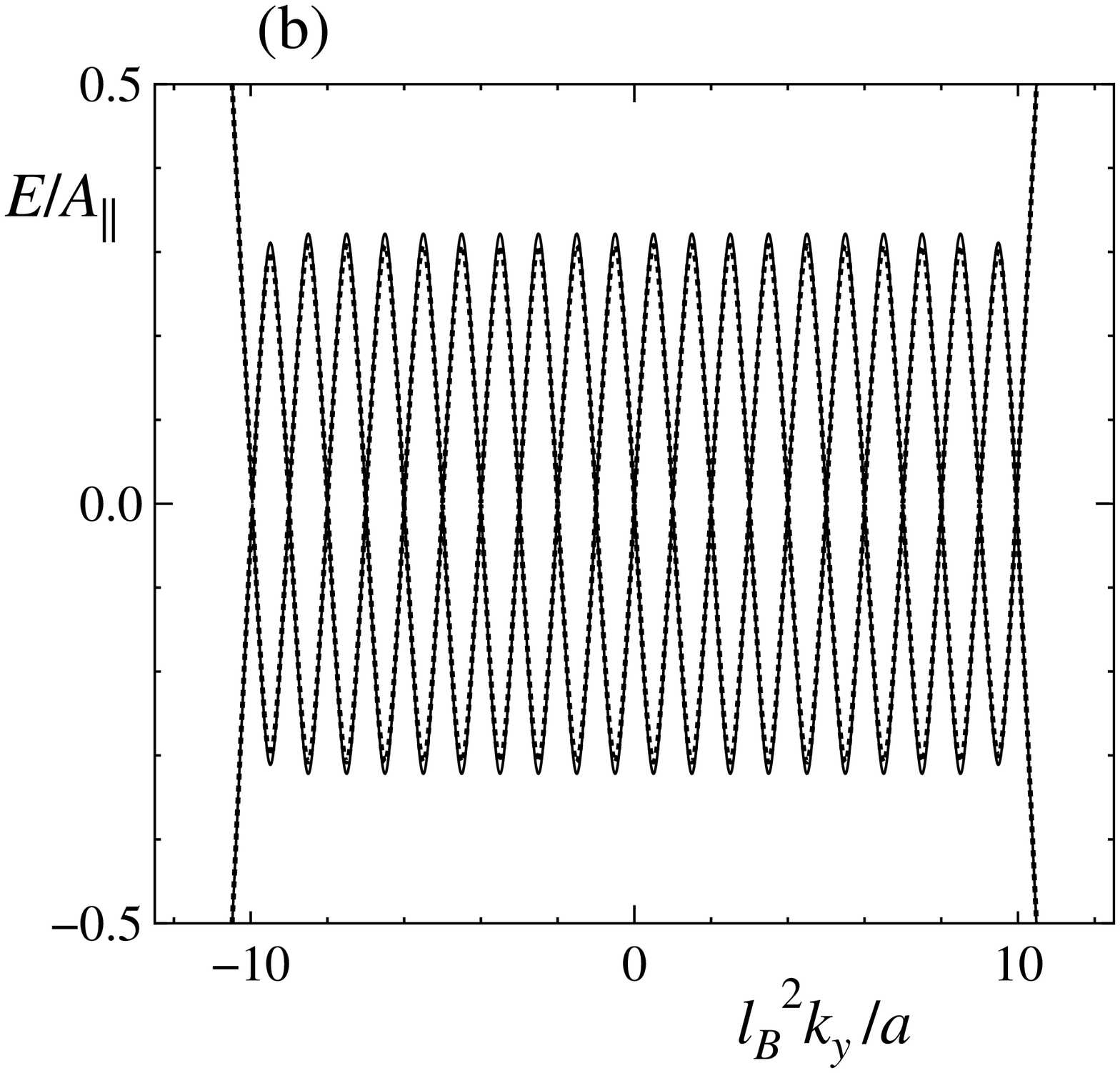}
\end{center}
\caption{
Band structures in the cases of (a) $l_{B}/a = 2$ and (b) $l_{B}/a = 1$
with $r = 0.4$ and $M = 21$,
where solid and dotted lines, respectively representing
the results of the 2D and 3D models,
almost completely overlap with each other.
}
\end{figure}
%%%%%%%%%%%%%%%%%%
One may think that the 2D model becomes unreliable with decreasing $l_{B}$
and the oscillatory behavior shown above is an artifact
induced by an erroneous application of the 2D model.
In the remainder of this section,
we show that this hypothesis can be ruled out.
To show this, we calculate the band structure in the case of
much stronger magnetic fields by using both the 2D model and the 3D model.
In determining the band structure on the basis of $H_{\rm 3D}$,
we consider an infinitely long rectangular prism-shaped system
of height $M$ and width $N$ (see Fig.~1).
Setting $M = 21$ and $N = 20$, we numerically determine the band structure
with the following parameters:
$A_{\perp}/A_{||} = 0.4$, $m_0/A_{||} = -0.6$, $m_{2\parallel}/A_{||} = 0.5$,
$m_{2\perp}/A_{||} = -0.1$.
Figure~5 shows the band structures obtained from the 2D and 3D models
in the cases of $l_{B}/a = 2$ and $1$.
The results are almost identical with each other.
This supports the reliability of $H_{\rm 2D}$ even in a strong magnetic field.

\section{Beyond the Continuum Dirac Theory}

In the previous section,
we observed the nontrivial behavior of the Landau level
that cannot be explained within the continuum Dirac theory.
We can consider that this reflects the bare character of 1D helical channels
constituting surface states of WTIs, as explained below.

As a transparent example, let us focus on the subband corresponding to
the $0$th Landau level.
If the mutual coupling of 1D helical channels is ignored, the energy of
the $j$th channel disconnected from neighboring ones is
\begin{align}
  \epsilon_{j \sigma}
  = \sigma v \left( k_{y}-\frac{z_{j}}{l_{B}^{2}} \right)
\end{align}
under the choice of $\mib{A}$ given in Eq.~(\ref{eq:def-A}),
where $\sigma$ = $+$ for the spin-up state and $\sigma = -$
for the spin-down state.
Hence, at
\begin{align}
   \frac{l_{B}^{2}k_{y}}{a} = \frac{z_{j}}{a} = j-\frac{M-1}{2} ,
\end{align}
the energy of the $j$th channel is $\epsilon_{j \sigma} = 0$
regardless of its spin state
while that of the nearest-neighbor channels is
$\epsilon_{j \pm 1 \sigma} = \mp \sigma va/l_{B}^{2}$.
Thus, we see that the coupling of the $j$th helical channel to
the $j+1$th and $j-1$th channels becomes weak with decreasing $l_{B}/a$.
This indicates that, in the vicinity of $l_{B}^{2}k_{y}/a = z_{j}/a$,
the subband state is governed by the $j$th channel
if $l_{B}/a$ is sufficiently small,
resulting in nearly linear energy dispersions.
This behavior is most clearly observed in the case of $l_{B}/a = 1$
shown in Fig.~5(b).
Generally speaking, the coupling of 1D helical channels is effectively
weakened by an applied magnetic field and, as a result, the character of
nearly disconnected 1D helical channels manifests itself
when $l_{B}/a$ is sufficiently small.

The explanation given above is based on an approach
from the strong magnetic field limit.
An alternative approach from the weak magnetic field limit is also possible.
We show below that the nontrivial behavior observed in the previous section
is explained within the continuum Dirac theory
by adding a correction that induces the mixing of the two Dirac cones.
Our attention is focused on surface states located near the center of
the system, and thus chiral edge states are beyond our consideration below.

We start with the approximate solutions of the eigenvalue equation
for $H_{\rm 2D}$, defined by
\begin{align}
   |\phi^{+}_{\pm n} \rangle
  & = \sqrt{a} \sum_{j} |j\rangle \mib{\phi}^{+}_{\pm n}(z_{j}) ,
       \\
   |\phi^{-}_{\pm n} \rangle
  & = \sqrt{a} \sum_{j} |j\rangle e^{i \pi j}\mib{\phi}^{-}_{\pm n}(z_{j}) ,
\end{align}
where the factor $e^{i \pi j}$ is attached to $|\phi^{-}_{\pm n} \rangle$
since it belongs to the second Dirac cone centered at
$(k_{y}, k_{z}) = (0, \pi/a)$.
In the limit of $l_{B} \gg a$, they satisfy
$H_{\rm 2D} |\phi^{+}_{\pm n} \rangle = E_{\pm n} |\phi^{+}_{\pm n} \rangle$
as well as the orthonormalization condition of
$\langle\phi^{\alpha}_{\pm n}|\phi^{\alpha^{\prime}}_{\pm n^{\prime}}\rangle
= \delta_{\alpha,\alpha^{\prime}}\delta_{n,n^{\prime}}$
with $\alpha$, $\alpha^{\prime} = +$ or $-$.
However, with decreasing $l_{B}$ toward $a$,
these equations are satisfied only in an approximate sense. 
Note that the term representing electron transfer between neighboring chains
in $H_{\rm 2D}$
is approximated by the derivative with respect to $z$ in the Dirac theory.
This approximation becomes worse with decreasing $l_{B}$.
Let us consider a correction for this approximation.

For definiteness, we consider the $n$th Landau level with $E_{n} \ge 0$
consisting of $|\phi^{+}_{n} \rangle$ in the first Dirac cone
and $|\phi^{-}_{n} \rangle$ in the second Dirac cone,
and evaluate the energy deviation from $E_{n}$ by taking account of
only the hybridization between them.
Since $\mib{\phi}^{\pm}_{n}(z)$ is the solution of the Dirac equation,
$|\phi^{\pm}_{n} \rangle$ satisfies
\begin{align}
      \label{eq:EE-correction}
   \left(H_{\rm 2D}-E_{n}\right)|\phi^{\pm}_{n} \rangle
 & = \pm i\frac{v^{\prime}}{2a}\sqrt{a}\sum_{j}|j\rangle \sigma_{y}
     P_{j}^{\pm}
       \nonumber \\
 & \hspace{-20mm}
     \times
     \left[ 2a\partial_{z}\mib{\phi}^{\pm}_{n}(z_{j})
            -\mib{\phi}^{\pm}_{n}(z_{j+1}) + \mib{\phi}^{\pm}_{n}(z_{j-1})
     \right] ,
\end{align}
where $P_{j}^{+} = 1$ and $P_{j}^{-} = e^{i \pi j}$.
The right-hand side of Eq.~(\ref{eq:EE-correction}),
including the difference between the derivative and the finite difference,
represents the error of the continuum approximation.
If $\langle\phi^{\alpha}_{n}|\phi^{\alpha^{\prime}}_{n}\rangle
= \delta_{\alpha,\alpha^{\prime}}$ is assumed, we obtain
\begin{align}
      \label{eq:correction_D}
    \langle\phi^{\pm}_{n}|H_{\rm 2D}|\phi^{\pm}_{n}\rangle
  & = E_{n} + \delta_{n} ,
          \\
      \label{eq:correction_OD}
    \langle\phi^{\pm}_{n}|H_{\rm 2D}|\phi^{\mp}_{n}\rangle
  & = \gamma_{n}
\end{align}
with
\begin{align}
   \delta_{n}
 & = \frac{v^{\prime}}{4}\sum_{j}
     \left( K_{j}^{n,n-1} - K_{j}^{n-1,n} \right) ,
         \\
   \gamma_{n}
 & = \frac{v^{\prime}}{4}\sum_{j}e^{i \pi j}
     \left( K_{j}^{n,n} - K_{j}^{n-1,n-1} \right)
\end{align}
for $n > 0$ and
\begin{align}
   \delta_{0}
 & = 0 ,
         \\
   \gamma_{0}
 & = \frac{v^{\prime}}{2}\sum_{j}e^{i \pi j} K_{j}^{0,0} ,
\end{align}
where
\begin{align}
   K_{j}^{n,m}
 = \psi_{n}(z_{j})
   \left[ 2a\partial_{z}\psi_{m}(z_{j})
          - \psi_{m}(z_{j+1}) + \psi_{m}(z_{j-1}) \right] .
\end{align}
$\delta_{n}$ and $\gamma_{n}$ appear to represent the correction
arising from the discreteness of the layered structure.
Note that they depend on $k_{y}$
as $\psi_{n}(z_{j})$ is centered at $z_{\rm c} = l_{B}^{2}k_{y}$.
Diagonalizing the $2 \times 2$ matrix composed of
Eqs.~(\ref{eq:correction_D}) and (\ref{eq:correction_OD}),
we find that the energy is shifted to
\begin{align}
      \label{eq:E-lowest-order}
  E_{n}^{\pm}(k_{y}) = E_{n} + \delta_{n}(k_{y}) \pm \gamma_{n}(k_{y}) .
\end{align}
In Fig.~6, Eq.~(\ref{eq:E-lowest-order}) is compared with the energy
of the $n$th Landau level obtained from $H_{\rm 2D}$
in the case of $l_{B}/a = 2.4$ for $n = 0$ and $1$.
We see that the modulation of the energy is roughly explained
within this treatment.
Since $\gamma_{n}$ as well as $\delta_{n}$ is proportional to
the difference between the derivative and the finite difference,
it becomes large with increasing $n$.
This accounts for the observation that the modulation becomes pronounced
with increasing index $n$ of the Landau level.
The absence of quantitative agreement implies that the coupling
between the Landau levels with different indices is also important.
%%%%%%%%%%%%%%%%%%
\begin{figure}[tbp]
\begin{center}
\includegraphics[height=4.0cm]{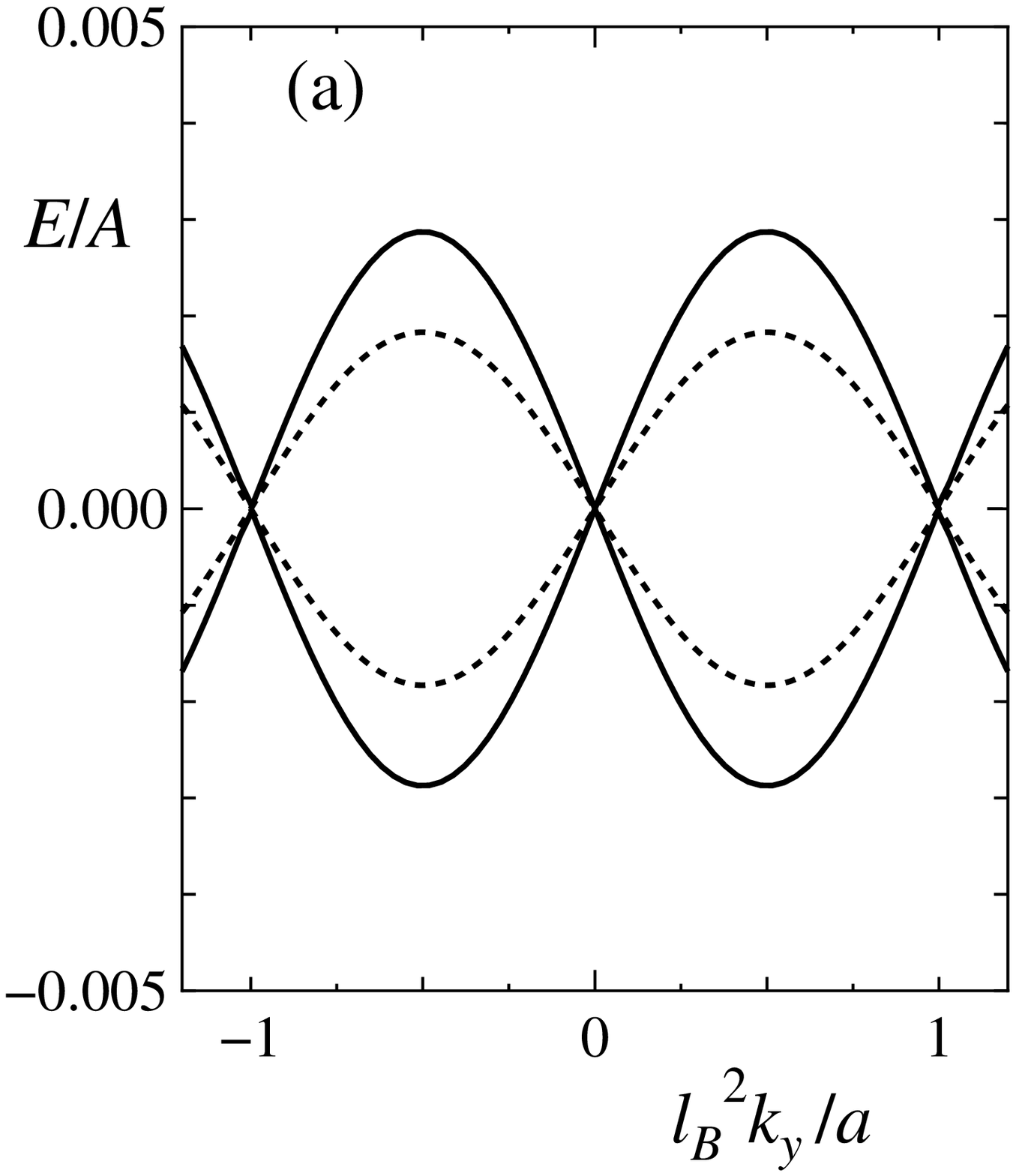}
\includegraphics[height=4.0cm]{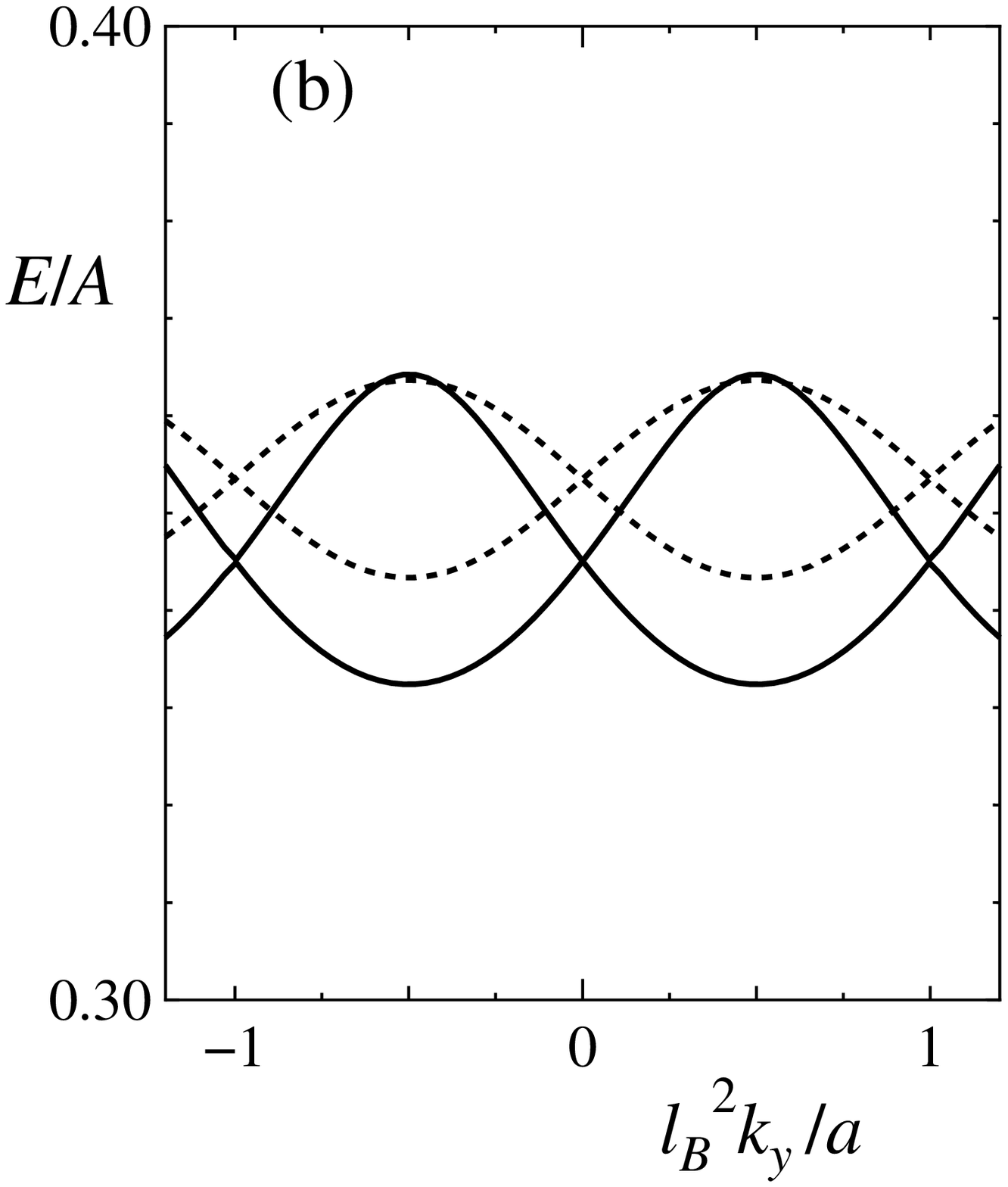}
\end{center}
\caption{
$k_{y}$ dependence of the energy of the $n$th Landau level
in the case of $l_{B}/a = 2.4$ for (a) $n = 0$ and (b) $n = 1$,
where solid lines represent the result obtained from $H_{\rm 2D}$
while dotted lines represent Eq.~(\ref{eq:E-lowest-order}).
}
\end{figure}
%%%%%%%%%%%%%%%%%%

The eigenstate corresponding to $E_{n}^{\pm}$ is obtained as
\begin{align}
  |\Phi^{\pm}_{n}\rangle
  = \sqrt{a} \sum_{j} |j\rangle \frac{1}{\sqrt{2}}
    \left[ \mib{\phi}^{+}_{n}(z_{j})
           \pm (-1)^{j} \mib{\phi}^{-}_{n}(z_{j}) \right] ,
\end{align}
which results in
\begin{align}
      \label{eq:Phi+_n}
   |\Phi^{+}_{n}\rangle
 & = \frac{\sqrt{a}}{2\sqrt{2}} \sum_{j} |j\rangle
       \nonumber \\
 & \hspace{-2mm} \times
     \left[ \begin{array}{c}
              \left(1+(-1)^{j}\right)
              \left(\psi_{n}(z_{j})-\psi_{n-1}(z_{j}) \right) \\
              \left(1-(-1)^{j}\right)
              \left(\psi_{n}(z_{j})+\psi_{n-1}(z_{j}) \right)
            \end{array}
     \right] ,
          \\
      \label{eq:Phi-_n}
   |\Phi^{-}_{n}\rangle
 & = \frac{\sqrt{a}}{2\sqrt{2}} \sum_{j} |j\rangle
       \nonumber \\
 & \hspace{-2mm} \times
     \left[ \begin{array}{c}
              \left(1-(-1)^{j}\right)
              \left(\psi_{n}(z_{j})-\psi_{n-1}(z_{j}) \right) \\
              \left(1+(-1)^{j}\right)
              \left(\psi_{n}(z_{j})+\psi_{n-1}(z_{j}) \right)
            \end{array}
     \right] 
\end{align}
for $n > 0$ and 
\begin{align}
      \label{eq:Phi+_0}
   |\Phi^{+}_{0}\rangle
 & = \frac{\sqrt{a}}{2} \sum_{j} |j\rangle
     \left[ \begin{array}{c}
              \left(1+(-1)^{j}\right)\psi_{0}(z_{j}) \\
              \left(1-(-1)^{j}\right)\psi_{0}(z_{j})
            \end{array}
     \right] ,
          \\
      \label{eq:Phi-_0}
   |\Phi^{-}_{0}\rangle
 & = \frac{\sqrt{a}}{2} \sum_{j} |j\rangle
     \left[ \begin{array}{c}
              \left(1-(-1)^{j}\right)\psi_{0}(z_{j}) \\
              \left(1+(-1)^{j}\right)\psi_{0}(z_{j})
            \end{array}
     \right] .
\end{align}
The above expressions indicate that in $|\Phi^{+}_{n}\rangle$,
the spin alternately polarizes to the $z$-direction in the chain with an even
$j$ and to the $-z$-direction in the chain with an odd $j$,
and the direction of spin polarization is completely reversed
in $|\Phi^{-}_{n}\rangle$.
Owing to this oscillatory behavior of the spin direction,
the spin state is not polarized
owing to self-averaging unless $l_{B}$ is very small.
This accounts for the absence of spin polarization
in the case of $l_{B}/a = 4$ [see Fig.~3(b)].
If $l_{B}$ decreases to a smaller value, self-averaging becomes
ineffective and the spin tends to polarize to the $+z$- or $-z$-direction
depending on whether the corresponding wave function is
dominant on the chains with an even $j$ or those with an odd $j$.
This accounts for the appearance of oscillatory spin polarization
in the case of $l_{B}/a = 2.4$ [see Fig.~4(b)].

The alternating behavior of $|\Phi^{\pm}_{n}\rangle$ is very natural
because $H_{\rm 2D}$ allows a spin-up (spin-down) 1D channel
to couple with only neighboring spin-down (spin-up) channels.
That is, each eigenstate of $H_{\rm 2D}$ should be represented
by the superposition of helical channels
with their spin direction alternately flipping from chain to chain.
This is precisely the feature represented by
Eqs.~(\ref{eq:Phi+_n})--(\ref{eq:Phi-_0}).

Let us consider the strong magnetic field regime where
$l_{B}/a$ is sufficiently small, focusing on the $0$th Landau level.
The spatial range of the superposition is on the order of $l_{B}$,
so $|\Phi^{\pm}_{0}\rangle$ is dominated by the channel
nearest to the center of the wave function.
This indicates that our argument based on the continuum Dirac theory is
continuously connected to the picture of nearly disconnected helical channels
argued in the beginning of this section.

\section{Summary and Discussion}

In this paper, we have studied how Dirac electrons on a surface of
weak topological insulators are quantized into Landau levels
in the presence of a perpendicular magnetic field $B$.
When $B$ is sufficiently weak with $l_{B}/a \gg 1$,
the ordinary Landau level structure with dispersion-less flat subbands
is observed in accordance with the quantization rule for Dirac electrons.
The spin state of each level is not polarized except for chiral edge states
appearing near the edges.
With increasing $B$, as $l_{B}/a \to 1$, each flat subband becomes oscillating
as a function of $l_{B}^{2}k_{y}/a$ with period $1$.
The corresponding spin state also becomes oscillating with the same period.
We have shown that this nontrivial behavior is attributed to
the mixing of the two Dirac cones induced by
the discreteness of the layered structure.
Alternatively, it can be regarded as a manifestation of the character of
a 1D helical channel nearly disconnected from neighboring ones
by a magnetic field.
We have also shown that these two explanations are consistent with each other.

Here, we discuss the possibility of detecting the nontrivial behavior
of the Landau level in actual experimental situations.
As its simplest consequence, we expect the broadening of each Landau level,
which will be detectable if its width is sufficiently large.
When a magnetic field of $B$ [T] is applied,
the magnetic length is expressed as $l_{B} = 25.65/\sqrt{B}$ [nm].
While the magnetic field of $B = 40$ T yields $l_{B} \approx 4.06$ nm,
the interlayer distance of the possible weak topological insulator
$\rm KHgSb$~\cite{yan} is $a \approx 1.9$ nm.
These parameters yield $l_{B}/a \approx 2.1$.
From Fig.~2(c) for the case of $l_{B}/a = 2.4$ and Fig.~5(a)
for the case of $l_{B}/a = 2$, we see that this value would result in
sufficiently large level broadening for its detection.
Similar broadening would also appear
in monolayer graphene.~\cite{wakabayashi,hasegawa,hatsugai}
However, an extremely strong magnetic field is necessary to detect it
as the lattice constant of graphene is on the order of 1 \AA.

In the presence of disorder, a further consideration is needed
as disorder also induces level broadening.
Roughly speaking, the broadening due to disorder is proportional
to $l_{B}^{-1} \propto \sqrt{B}$
and is independent of the index $n$.~\cite{koshino}
Hence, the broadening due to disorder slowly increases with increasing $B$,
while the oscillation-induced broadening rapidly increases
in the strong magnetic field regime, as can be seen from Figs.~2 and 5.
Furthermore, the former is independent of the index $n$
while the latter becomes large with increasing $n$.
We expect that these qualitative differences enable us to detect
the oscillation-induced broadening as long as the disorder is not so strong.

So far, the Zeeman effect on Landau levels has been ignored in our argument.
Here, we briefly consider it within the continuum Dirac theory.
If we take into account the Zeeman term
for a magnetic field $\mib{B}=(B,0,0)$, the Hamiltonian $H_{\pm}$ becomes
\begin{align}
   \tilde{H}_{\pm}
   = \left[ 
       \begin{array}{cc}
         v(k_{y}-eBz) & \mp v^{\prime}\partial_{z} + \Delta \\
         \pm v^{\prime}\partial_{z} + \Delta & -v(k_{y}-eBz)
       \end{array}
     \right]
\end{align}
with
\begin{align}
   \Delta = \frac{1}{2}\mu_{\rm B}g B ,
\end{align}
where $\mu_{\rm B}$ and $g$ are respectively the Bohr magneton and
the effective $g$ factor.
In the presence of the Zeeman term,
the energy of the $n$th Landau level is modified as
\begin{align}
  \tilde{E}_{\pm n}
  = \pm \frac{\sqrt{r}v}{l_B}
    \sqrt{2\left(n+\tilde{\Delta}^{2}\right)} ,
\end{align}
where
\begin{align}
  \tilde{\Delta} = \frac{l_{B}}{\sqrt{2r}v}\Delta .
\end{align}
If $v = 3$ eV\AA\ is assumed, we find that
$\tilde{\Delta} \approx 0.011 g/\sqrt{r}$ at $B = 40$ T.
The above argument implies that the Zeeman effect induces
only a small shift of each quantized level
and does not strongly alter the qualitative behavior of
the magnetic field effect on Landau levels.

Let us finally examine the stability of surface states
against an increase in the magnetic field.
For definiteness, we consider the $0$th Landau level
in the vicinity of $l_{B}^{2}k_{y}/a = z_{j}/a$.
The corresponding wave function is centered at the $j$th chain
and its spatial range is on the order of $l_{B}$.
A plausible criterion to ensure its stability is that all helical channels
within the distance of $l_{B}$ from the $j$th chain stably exist.
Equations~(\ref{eq:Mk-dis}) and (\ref{eq:Mk-dis><0})
indicate that this is ensured under the condition of
\begin{align}
       \label{eq:cond-suf}
   \tilde{m}_{0} + m_{2\parallel}\left(\frac{a}{l_{B}}\right)^{2} < 0
\end{align}
with $\tilde{m}_{0} = {\rm max}\{m_{0},m_{0}+4m_{2\perp}\}$,
where $m_{2\parallel} > 0 > \tilde{m}_{0}$ is assumed.
It is convenient to define the characteristic length $l_{B}^{\rm c}$
as $l_{B}^{\rm c}/a \equiv (m_{2\parallel}/|\tilde{m}_{0}|)^{1/2}$.
If $l_{B} > l_{B}^{\rm c}$, condition~(\ref{eq:cond-suf}) is satisfied,
resulting in the stabilization of the surface state.
However, this does not mean that the surface state completely disappears
when $l_{B}^{\rm c} > l_{B}$.
Even in this regime, the helical channel in the $j$th chain persists
and its energy is zero at $l_{B}^{2}k_{y}/a = z_{j}/a$,
but helical channels in its neighboring chains
within the distance of $l_{B}$ are partly or fully destabilized and are
replaced with bulk states.
As a consequence, the persistent $j$th helical channel is inevitably coupled
with such bulk states.
Hence, although the resulting state contains the $j$th helical channel,
it cannot be regarded as a pure surface state.

\section*{Acknowledgment}

The author thanks T. Arita for technical assistance in numerical computations.
This work was supported by a Grant-in-Aid for Scientific Research (C)
(No. 15K05130).


\begin{thebibliography}{99}

\bibitem{fu} L. Fu, C. L. Kane, and E. J. Mele,
Phys. Rev. Lett. {\bf 98}, 106803 (2007).

\bibitem{moore} J. E. Moore and L. Balents,
Phys. Rev. B {\bf 75}, 121306 (2007).

\bibitem{roy} R. Roy, Phys. Rev. B {\bf 79}, 195322 (2009).

\bibitem{kane} C. L. Kane and E. J. Mele,
Phys. Rev. Lett. {\bf 95}, 146802 (2005).

\bibitem{bernevig} B. A. Bernevig and S.-C. Zhang,
Phys. Rev. Lett. {\bf 96}, 106802 (2006).

\bibitem{ran} Y. Ran, Y. Zhang, and A. Vishwanath,
Nat. Phys. {\bf 5}, 298 (2009).

\bibitem{imura1} K.-I. Imura, Y. Takane, and A. Tanaka,
Phys. Rev. B {\bf 84}, 195406 (2011).

\bibitem{ringel} Z. Ringel, Y. E. Kraus, and A. Stern,
Phys. Rev. B {\bf 86}, 045102 (2012).

\bibitem{mong} R. S. K. Mong, J. H. Bardarson, and J. E. Moore,
Phys. Rev. Lett. {\bf 108}, 076804 (2012).

\bibitem{liu1} C.-X. Liu, X.-L. Qi, and S.-C. Zhang,
Physica E {\bf 44}, 906 (2012).

\bibitem{imura2} K.-I. Imura, M. Okamoto, Y. Yoshimura, Y. Takane,
and T. Ohtsuki, Phys. Rev. B {\bf 86}, 245436 (2012).

\bibitem{yoshimura} Y. Yoshimura, A. Matsumoto, Y. Takane, and K.-I. Imura,
Phys. Rev. B {\bf 88}, 045408 (2013).

\bibitem{kobayashi} K. Kobayashi, T. Ohtsuki, and K.-I. Imura,
Phys. Rev. Lett. {\bf 110}, 236803 (2013).

\bibitem{morimoto1} T. Morimoto and A. Furusaki,
Phys. Rev. B {\bf 89}, 035117 (2014).

\bibitem{obuse} H. Obuse, S. Ryu, A. Furusaki, and C. Mudry,
Phys. Rev. B {\bf 89}, 155315 (2014).

\bibitem{takane} Y. Takane, J. Phys. Soc. Jpn. {\bf 83}, 103706 (2014).

\bibitem{yan} B.-H. Yan, L. M\"{u}chler, and C. Felser,
Phys. Rev. Lett. {\bf 109}, 116406 (2012).

\bibitem{rasche} B. Rasche, A. Isaeva, M. Ruck, S. Borisenko, V. Zabolotnyy,
B. Buchner, K. Koepernik, C. Ortix, M. Richter, and J. van den Brink,
Nat. Mater. {\bf 12}, 422 (2013). 

\bibitem{tang} P. Tang, B. Yan, W. Cao, S.-C. Wu, C. Felser, and W. Duan,
Phys. Rev. B {\bf 89}, 041409 (2014).

\bibitem{g-yang} G. Yang, J. Liu, L. Fu, W. Duan, and C. Liu,
Phys. Rev. B {\bf 89}, 085312 (2014). 

\bibitem{pauly} C. Pauly, B. Rasche, K. Koepernik, M. Liebmann, M. Pratzer,
M. Richter, J. Kellner, M. Eschbach, B. Kaufmann, L. Plucinski,
C. M. Schneider, M. Ruck, J. van den Brink, and M. Morgenstern,
Nat. Phys. {\bf 11}, 338 (2015).

\bibitem{shon} N. H. Shon and T. Ando,
J. Phys. Soc. Jpn. {\bf 67}, 2421 (1998).

\bibitem{cheng} P. Cheng, C. Song, T. Zhang, Y. Zhang, Y. Wang, J.-F. Jia,
J. Wang, Y. Wang, B.-F. Zhu, X. Chen, X. Ma, K. He, L. Wang, X. Dai, Z. Fang,
X. Xie, X.-L. Qi, C.-X. Liu, S.-C. Zhang, and Q.-K. Xue,
Phys. Rev. Lett. {\bf 105}, 076801 (2010).

\bibitem{hanaguri} T. Hanaguri, K. Igarashi, M. Kawamura, H. Takagi,
and T. Sasagawa, Phys. Rev. B {\bf 82}, 081305 (2010).

\bibitem{novoselov} K. S. Novoselov, A. K. Geim, S. V. Morozov, D. Jiang,
M. I. Katsnelson, I. V. Grigorieva, S. V. Dubonos, and A. A. Firsov,
Nature {\bf 438}, 197 (2005).

\bibitem{liu2} C.-X. Liu, X.-L. Qi, H. Zhang, X. Dai, Z. Fang,
and S.-C. Zhang, Phys. Rev. B {\bf 82}, 045122 (2010).

\bibitem{z-yang} Z. Yang and J.-H. Han,
Phys. Rev. B {\bf 83}, 045415 (2011).

\bibitem{ilan} R. Ilan, F. de Juan, and J. E. Moore, arXiv:1410.5823.

\bibitem{morimoto2} T. Morimoto, A. Furusaki, and N. Nagaosa,
Phys. Rev. Lett. {\bf 114}, 146803 (2015).

\bibitem{arita} T. Arita and Y. Takane,
J. Phys. Soc. Jpn. {\bf 83}, 124716 (2014).

\bibitem{comment} More precisely speaking,
the spatial range of $\psi_{n}(z)$ is on the order of $\sqrt{n+1/2}l_{B}$,
so the $n$th Landau level stably exists under the condition of
$(M-1)/2-\sqrt{n+1/2}l_{B}/a \gtrsim l_{B}^{2}|k_{y}|/a$.

\bibitem{wakabayashi} K. Wakabayashi, M. Fujita, H. Ajiki, and M. Sigrist,
Phys. Rev. B {\bf 59}, 8271 (1999).

\bibitem{hasegawa} Y. Hasegawa and M. Kohmoto,
Phys. Rev. B {\bf 74}, 155415 (2006).

\bibitem{hatsugai} Y. Hatsugai, T. Fukui, and H. Aoki,
Phys. Rev. B {\bf 74}, 205414 (2006).

\bibitem{koshino} M. Koshino and T. Ando,
Phys. Rev. B {\bf 75}, 235333 (2007).



\end{thebibliography}
\end{document}